\documentclass[prb,aps,twocolumn,amsmath,showpacs]{revtex4}
\usepackage{graphicx}
\usepackage[]{epstopdf}
\usepackage[utf8]{inputenc}
\usepackage{color}
\begin{document}
\title{Current rectification in molecular junctions produced
by local potential fields}
\author{Tomasz Kostyrko}
\affiliation{Faculty of Physics, A. Mickiewicz University,
ul.~Umultowska~85,61-614~Pozna\'{n},~Poland}

\author{V\'{\i}ctor M. Garc\'{\i}a-Su\'arez}
\affiliation{Departamento de F\'{\i}sica, Universidad de Oviedo \&
CINN, 33007 Oviedo, Spain}
\affiliation{Department of Physics, Lancaster University,
Lancaster, LA1 4YB, United Kingdom}

\author{Colin J. Lambert}
\affiliation{Department of Physics, Lancaster University,
Lancaster, LA1 4YB, United Kingdom}

\author{Bogdan R. Bu{\l}ka}
\affiliation{Institute of Molecular Physics, Polish Academy of
Science, Smoluchowskiego 17, 60-179 Pozna{\'n}, Poland}
\begin{abstract}
The transport properties of a octane-dithiol (ODT) molecule
coupled to Au(001) leads are analyzed using density functional
theory and non-equilibrium Green functions. It is shown that a
symmetric molecule can turn into a diode under influence of a
local electric field created by an external charged probe. The
origin of the asymmetry of the current\---voltage ($I-V$)
dependence is traced back to the appearance of a probe induced
quasi\---local state in the pseudogap of the ODT molecule. The
induced state affects electron transport, provided
it is close to the Fermi level of the leads. An asymmetric
placement of the charged probe along the alkane chain makes the
induced quasi\---local state in the energy gap very sensitive to
the bias voltage and results in rectification of the current. The
results based on DFT are supported by independent calculations
using a simple one\---particle model Hamiltonian.
\end{abstract}
\pacs{73.63.-b, 85.65.+h, 73.40.-c}
\date{\today, revised version}

\maketitle
\section{Introduction}
The quest for molecular diodes, which was started by Aviram and Ratner
in 1974 \cite{Aviram-74} has stimulated a number of experimental and
theoretical investigations of molecular rectifiers\cite{Ashwell-93,
  Ellenbogen-00, Krzeminski-01, Kornilovitch-02, Taylor-03, Larade-03,
  Troisi-04, Elbing-05, Metzger-06, Liu-06, Dalgleish-06,
  Armstrong-07, Stadler-08, Ford-08, Diez-Perez-09}. The
  Aviram-Ratner mechanism is based on donor-insulator-acceptor
  molecules, where electrons are transferred inelastically from the
  acceptor to the donor for a certain bias value and polarity.  It is
  also possible to have rectifying behaviours in case of coherent
  transport when asymmetries are present in the molecular
  structure. Such asymmetries can be produced for instance by
  asymmetric molecules or different contacts to the electrodes.  The
  current starts to increase for a certain bias and polarity whenever
  molecular orbitals which are located asymmetrically with respect to
  the leads coincide with the Fermi level of one of the electrodes
  \cite{Krzeminski-01, Kornilovitch-02, Larade-03} or when the
  energies of molecular orbitals localized on different parts of the
  molecule match inside the bias window. For the reverse polarity the
  orbitals located asymmetrically reach one of the Fermi levels at
  higher voltages, in the first case, and they separate in energy and
  do not match inside the bias window, in the second case. The coherent
  mechanism gives rise to moderately high rectification ratios,
  $RR(V)=|I(V)/I(-V)|$ (where $I(V)$ denotes the current as a function
  of the voltage $V$).\cite{Ellenbogen-00} The theoretical
calculations based on either coherent or incoherent mechanisms
predict clear rectification behaviors but from a practical point of
view success has been modest.\cite{Ashwell-93, Zhou-97, Yao-99}
Estimates using both parametric models and {\it ab initio} studies
have raised doubts about whether the molecular asymmetries could lead
to a sizable $RR$ based on elastic scattering.\cite{Kornilovitch-02,
  Taylor-03, Larade-03, Troisi-04, Liu-06, Dalgleish-06, Stadler-08,
  Ford-08} Most of these studies report values of $RR$ which do not
exceed a dozen or so.  So far however the question whether there is an
upper limit for the rectification value for a molecular diode, and
what this value might be, did not receive an unambiguous answer.

The experimentally obtained values of $RR$ for molecular diodes are
usually much less than $100$, and are orders of magnitude smaller than
the corresponding values in typical inorganic devices.\cite{Elbing-05,
  Metzger-06, Diez-Perez-09} Although much bigger values
($RR\sim{}10^3$) of $RR$ may be obtained after exposure of the
self\---assembled monolayers to air\cite{McCreery-04} or by
protonation\cite{Ashwell-04}, it is not clear the possible impact of
the ionic conductivity on the current\---voltage characteristics of
the samples.\cite{Metzger-08}

In order to design the best molecular diode it is necessary to
optimize the main parameters governing the rectifying behaviour of the
device: HOMO-LUMO energy gap (HOMO: for the highest occupied molecular
orbital, LUMO for the lowest unoccupied one), length of the
intervening molecular bridge, the position of the HOMO-LUMO system
with respect to the Fermi level of the leads, and the coupling
strength of the molecular levels to the metallic levels of the leads.
In both experimental and theoretical works this goal is commonly
realized by a more or less arbitrary choice of molecular components,
i.e. the donor and acceptor subunits and the connecting bridge.

In this paper, as an alternative approach to other molecular-scale
rectification mechanisms, we carry out a systematic simulation study
of rectification induced by the presence of a local external charge in
the vicinity of a molecular bridge. In part, our study is motivated by
related experiments involving carbon nanotubes with charged AFM tips
acting as the source of a local potential.\cite{Freitag-01} At a
molecular scale such a symmetry-breaking charge may be located in the
surrounding solvent or in a self\---assembled monolayer matrix of
shorter molecules with a substantial dipole moment. For our
calculations we use an octane\---dithiol molecule coupled to gold
leads, a system well studied in recent works \cite{Muller-06}.
\section{System setup and parameters of the DFT computations}
\begin{figure}[h]
\includegraphics*[width=0.47\columnwidth]{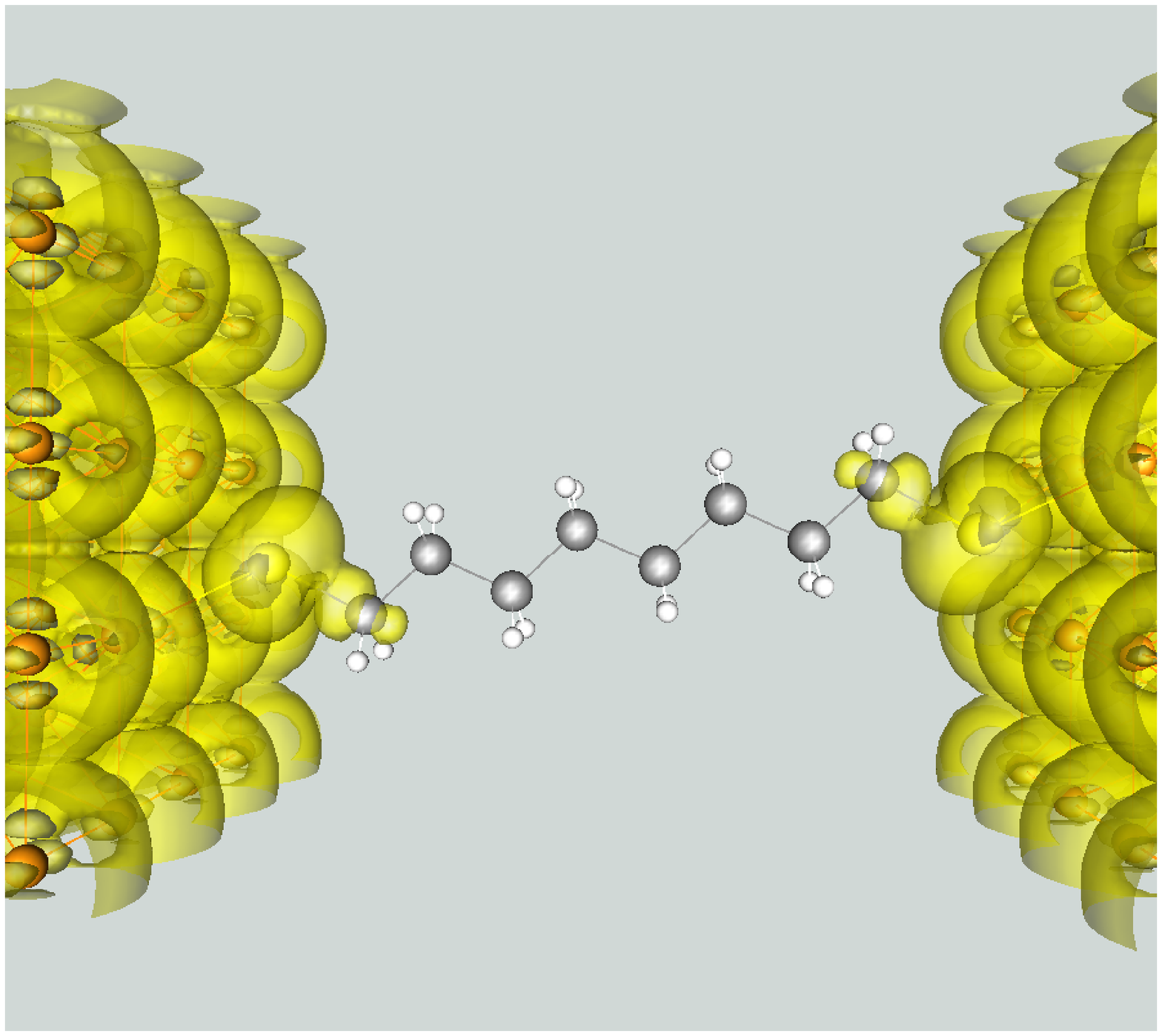}
\hspace*{0.04\columnwidth}\includegraphics*[width=0.47\columnwidth]{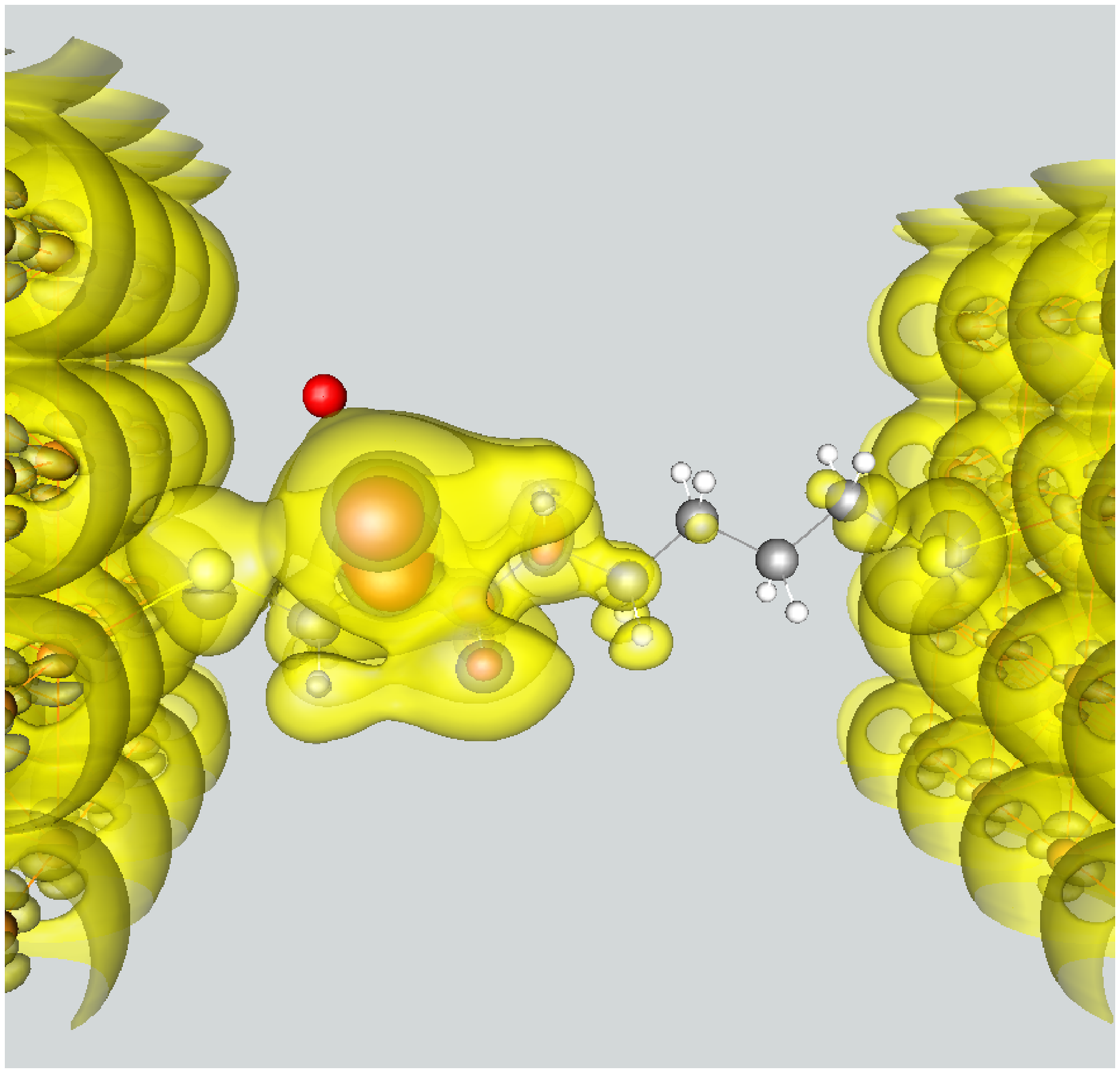}
\caption{\label{junction}(color online) The ball\---and\---stick model
  of the junction with the octane\---dithiol (ODT) molecule and
  fragments of the gold leads, used in our DFT computations. The
  regular balls indicate the positions of the atoms. The irregular
  contours in the figure show isosurfaces of local density of states
  (LDOS; see text in the end of Section III for a discussion) near the
  energy $E_F+1.2$~eV, integrated in the energy window of width
  1~eV. The left figure is for the junction without the probe and the
  right one is for the junction with the probe in position C1. The
  probe is shown with an additional (red) ball in the right figure.}
\end{figure}

We model the molecular junction within the {\sc smeagol}
methodology\cite{smeagol}, where the system is divided into an
extended molecule (formed by the molecule and some surface layers of
gold atoms) and the periodic, semi-infinite leads. In our case the
extended molecule is the ODT molecule and 4$\times$4$\times$4 and
4$\times$4$\times$5 atomic sections of Au FCC lattice on the left and
right side of the junction, respectively (see figure~\ref{junction}).
Periodic boundary conditions are applied in
the directions parallel to the planes, i.e. transverse to the
conduction direction. The solutions at the right and
left boundary of the extended molecule are matched to the
solution obtained from an independent computation of the
periodic leads.  The total number of atoms in the unit cell was
170. The spatial structure of the junction was defined in several
stages. First, the structure of the periodic leads and the ODT
molecule was independently optimized using a conjugate gradient
method from {\sc siesta},\cite{siesta} with a maximum force
tolerance of 0.01~eV/\AA. Next, the ODT molecule was attached in
the hollow site position of the Au(001) surface of the leads and
the molecular structure was optimized again (we did not allow at
this stage for the relaxation of Au atoms) with the same
parameters. In the course of the relaxation the bonding S atoms on
both sides moved significantly to the bridge position of the
Au(001) surface plane, with dominating 2-Au coordination.

In the {\sc siesta} computations\cite{siesta-a} we used the local
density approximation (LDA) with the Perdew and Zunger\cite{Perdew-81}
model for the exchange and correlation potential. The core electrons
were represented by pseudopotentials obtained using the Troullier and
Martins method.\cite{Troullier-91} For all atoms in the system we
employed a single-zeta (SZ) basis set. This restriction was necessary
due to the memory limitations of our computation facilities,
especially in the subsequent {\sc smeagol} transport computations.
However, tests performed with smaller gold leads and a double-zeta
polarized basis set (DZP) or the same gold leads and double-zeta (DZ)
or single-zeta polarized (SZP) basis sets showed the same qualitative trends
in the transport properties and the electronic structure (density
of states (DOS)). The radii of the pseudoatomic orbitals were
fixed by an energy shift parameter of 0.02~Ry. The mesh cutoff
parameter was set to 200~Ry, and we used an additional grid cell
sampling.

To simulate the effect of the charge probe we used an alkali atom
(charge $+|e|$) or an alkaline earth atom (charge $+2|e|$) with the
basis restricted to just the $s$ orbital and a very small cutoff
radius. This trick raises the atomic level of the probe atom far above
Fermi level of the leads, completely depleting the $s$ orbital. In
effect, the auxiliary atom can be considered to be a simple point
charge, whose wave function has no overlap with the wave functions of
ODT and gold. The unit cell remained exactly neutral in the {\sc
  siesta} computation, and was approximately neutral in the {\sc
  smeagol} computation. The atomic forces, upon introducing the probe
atom, turned out to be quite large due to Coulomb attraction between
the displaced electronic charges and the ionic charge of the probe
center. However, we did not re\---optimize the spatial structure of
the unit cell at this stage, since the charge probe is meant to be an
abstract representation of a local potential which could have various
origins.
\section{Results}
\begin{figure}[h]
\includegraphics*[width=\columnwidth]{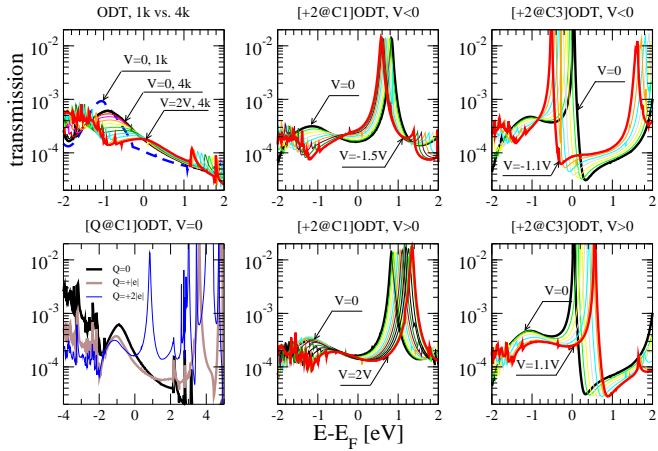}
\caption{\label{transmission_ODT}(color online) Evolution of the
  transmission versus energy near the Fermi level ($E_{\mathrm
    F}\equiv0$) as a function of the bias voltage for the
    different values and the positions of the probe charge, obtained
    from {\sc smeagol} for the junction with ODT. All the plotted
    curves but one correspond to $4k$ computations.  Left column,
    upper row: the junction without the probe, $4k$ results for
    0$<V<$2V are compared with $1k$ for $V=0$. Left column, lower row:
    transmission function for various values of the charge probe in
    the extended energy scale. Middle column: the probe is positioned
    near the leftmost C atom (position C1, as in
    figure~\protect\ref{junction}) of the ODT chain at a distance
    ca$.$ 4~\AA\/ from the carbon atom ($V<0$: upper row, $V>0$: lower
    row). Right column: the probe is shifted to the position of third
    C atom from the left, C3 ($V<0$: upper row, $V>0$: lower row). The
    heavy lines show the results for the limiting values of voltage
    indicated with arrows, and the thin ones are for intermediate
    values of voltage.}
\end{figure}

Figure~\ref{transmission_ODT} shows the computed electron transmission
coefficients, obtained for both 1$k$-point and
4$k$-points\cite{siesta-a} in the transverse direction of the
Brillouin zone. Since the large transverse size of the unit cell
produces a large separation between different copies of the molecular
junction, the results are a representative of transport through an
individual molecule.  Furthermore, a sufficiently-large transverse
unit cell needs only a small number of transverse $k$-points to
approach the limit of the infinite surface. In our case, however we
can see that the difference between 1$k$ and 4$k$-point calculation is
still noticeable, although the main features of the results remain
qualitatively the same. As shown in figure~\ref{transmission_ODT}, a
wide (ca$.$ 2~eV) transmission peak is located at about 1~eV below the
Fermi level of the system. The peak is broad because it comes from
molecular orbitals which are well coupled to the electronic states of
the leads, i.e. the sulfur $3s$ and especially $3p$ orbitals. Note,
that the maximum height of the peak is much less than unity, because
it arises from two almost energetically equivalent S orbitals at
opposite sites of the junction \cite{Kostyrko-02,lazslo} with an
effective coupling, which decreases exponentially with the length of
the molecular backbone. Under a bias voltage the peak splits into two
peaks, with one moving upward and the other downward in energy. This
behaviour is due to the influence of the potential ramp of the applied
voltage, which induces an energetic inequivalence of the S orbitals at
opposite ends of the molecule.

Influence of the charge of the probe on the zero\---voltage
  transmission is presented in the lower left panel of
  figure~\ref{transmission_ODT}. In the vicinity of the Fermi level
  the $Q=+|e|$ charge merely produces a rather uniform suppression of
  the of the wide transmission peak located about 1~eV below
  $E_{\mathrm F}$.  More important is appearance of a rather prominent
  narrow transmission peak located at $E\sim3.6$~eV above $E_{\mathrm
    F}$, and which does not have a counterpart in the transmission
  without the probe. For the $Q=+|e|$ charge probe the peak will not
  manifest itself in the corresponding I-V dependence, because it is
  too far from the Fermi level and would not enter the source\---drain
  voltage window for moderate ($|V|<$2V) voltage values.  However
  increase of the charge of the probe to a value of $+2|e|$ appears to
  shift this peak to within a distance of $\sim1$~eV from
  $E_{\mathrm{}F}$, which makes it important for the transport for
  $|V|<$2V.

 By comparing the middle and the right column of figure
  \ref{transmission_ODT} one can note a significant dependence of the
  transmission peak on the position of the charge probe.  Shifting the
  probe from the left side (position C1) of the chain to a position
  closer to the center (C3) increases the height of the
  zero\---voltage peak and moves it close to the Fermi level of the
  system.

\begin{figure}[h]
\includegraphics*[width=\columnwidth]{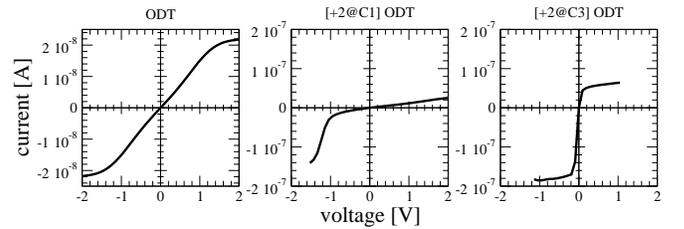}
\caption{\label{current_ODT} $I-V$ dependence for the ODT molecular
  junction. Left figure: without the charge probe. Middle figure: the
  $+2|e|$ charge probe is near the leftmost C atom (C1). Right figure:
  the $+2|e|$ charge probe is near the third C atom from the left
  (C3).}
\end{figure}

The evolution of the transmission with the bias voltage is
  presented in the middle and right columns of
  figure~\ref{transmission_ODT} for two positions of the probe. For
negative bias voltages (i.e. when the electrode nearest the probe
charge is negative) the narrow peak moves in the direction of
decreasing energy values, i.e. it approaches the Fermi level as the
absolute value of the bias voltage increases.  For positive bias
voltages this behaviour reverses. In the case of negative bias
voltages the peak enters the transport window for some threshold
voltage value, while it escapes the voltage window for positive
biases.  As shown in figure \ref{current_ODT}, this behaviour leads to
an asymmetric I-V characteristics and a significant rectification
ratio. (note the changes of scale in the vertical
axes).\cite{convergence}
For the probe position C3 the rectification ratio reaches a maximum of
about 7, near the point where the sharp peak crosses the edge of the
transmission window. The asymmetry of the $I-V$ characteristic and the
rectification ratio both decrease when the probe is moved to the
center of the junction. At the same time, the maximum current in the
computed voltage range increases, consistent with the increased
  height of the peak for position C3.
  
\begin{figure}[h]
\includegraphics*[width=\columnwidth]{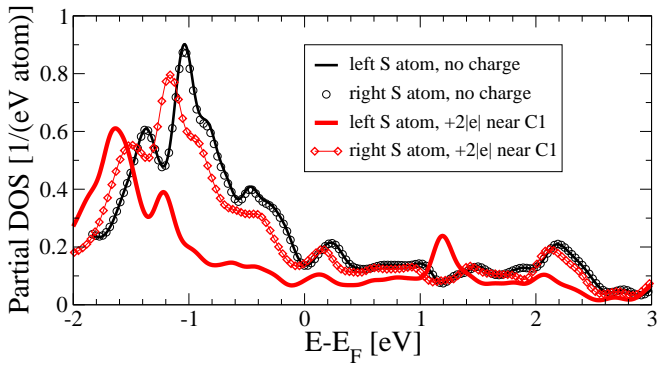}
\caption{\label{PDOS}(color online) Partial density of states (PDOS,
  summed over the sulfur valence orbitals) for the sulfur atoms of ODT
  molecule coupled to the leads, at $V=0$.}
\end{figure}

In order to understand the nature of the peak that appears in the
transmission for the stronger charge probe we computed the partial
density of states for the bonding S atoms when the charge probe is
located near the first C atom from the left of the ODT chain. The
results, presented in figure~\ref{PDOS}, were obtained using {\sc
  siesta} and mirror the transmission data of figure
\ref{transmission_ODT}, obtained from {\sc smeagol}.

As one can see there is indeed a wide (ca$.$ 2~eV) spectral
feature in the PDOS on both S atoms. Without the charge probe the
PDOSs on both S atoms are exactly the same, as expected for a
perfectly symmetric junction. The introduction of the charge probe
into the unit cell does not perturb significantly the local electronic
structure of the more distant S atom. However, the PDOS of the S atom
in the vicinity of the charge probe undergoes a substantial
change. First, the PDOS below $E_{\mathrm F}$ and especially the wide
band located near -1~eV are considerably reduced and shifted
downwards. This effect is in agreement with the observed increase of
the electronic charge on the S atom closer to the charge probe,
calculated from the Mulliken populations. Second, a rather narrow peak
emerges near 1~eV above $E_{\mathrm F}$, suggesting a close relation
with the pronounced peak at the same energy in the transmission
function. Since this peak is absent in the PDOS for the S atom in the
absence of the probe and for the more distant S atom, we conclude that
the most important effect of the charge probe is to introduce a narrow
peak in the transmission function and in the PDOS of the closer S
atom.

A small width of this peak suggests that it is related to the
appearance of a quasilocalized level induced in the pseudogap of the
ODT molecule coupled to the Au leads, and generated by the strong
local potential of the charge probe. The quasilocalized state, having
a finite spatial extent along the molecular chain, can couple to the
electronic states of both electrodes and in this way it can contribute
to the transmission. If the center of the gravity of the
quasilocalized state is closer to one of the electrodes, its energy
dependence on the voltage will be controlled by that electrode,
thereby inducing an asymmetry in the I-V characteristic.

To confirm the quasilocalized character of the probe\---induced
  peak we computed the local density of states (LDOS) from the
  vicinity of $E\sim{}E_F+1.2$~eV, where the peak is found in PDOS. The
  results are presented in figure~\ref{junction} for a junction without
  the probe (left panel) and the $+2|e|$ probe in position C1 (right
  panel).
 Without the probe LDOS is concentrated mainly near the S atoms
  and the neighboring C atoms with very little LDOS inside the alkane
  backbone. This is consistent with the approximate picture of a whole
  junction as a simple two\---S-atom system with a strong bonding to
  the leads and a weak bridge between them. Including the probe
  creates the considerable increase of LDOS near to the probe. Note,
  that although the biggest increase of LDOS takes place at the nearby
  C and H atoms, also the S atom closer to the probe gains noticeably
  more LDOS. At the same time, the more distant S atom is hardly
  influenced by the probe, consistent with what we found in PDOS.  On
  this basis we conclude that the probe induced changes are localized
  near the position of the probe.  Computations of LDOS for an energy
  window that does not include peak in PDOS (near $E_F+0.25$~eV, not
  shown here), exhibits much more symmetrical LDOS, more like to a
  system without the probe. The latter fact again confirms that the
  probe induced changes are limited to a rather narrow energy window
  and stresses the similarity of the peak to a quasi\---localized
  state in extended systems.\cite{Choi-00}

\section{Model interpretation of the DFT results}
To rationalize the above DFT results, we  now examine a simple
parametric tight binding model of the molecule, described by the
following Hamiltonian:
\begin{eqnarray}\label{H_M}
{\cal H}_{M} &=& t_{a}\sum_{i=1}^{N-1}\left(a^\dagger_ia_{i+1}+
\mbox{h.c.}\right) + E_a\sum_{i=1}^{N} a^\dagger_ia_i
+ \nonumber \\
&+& t_{b}\sum_{i=1}^{N-1}\left(b^\dagger_ib_{i+1}+ \mbox{h.c.}\right)+ E_b\sum_{i=1}^{N} b^\dagger_ib_i + \nonumber \\
&+& t_{ab}\sum_{i=1}^{N-1}\left(b^\dagger_ia_{i+1}+
a^\dagger_ib_{i+1}+\mbox{h.c.}\right)
\end{eqnarray}
In order to mimic the semiconducting properties of the ODT molecule it
includes two electron levels per site, one corresponding to the
bonding state and the other one to the antibonding state. These levels
are separated by an energy gap, $E_a-E_b>0$. The operators $a,b$ are
electron operators for the antibonding and the bonding levels. Note
that we suppressed here the spin index, since the only effect of the
spin in the absence of the magnetic field or electron repulsion
reduces to the appearance of a factor of 2 in the expression for the
current or the conductance.  We allow for hopping between nearest
neighbor sites only. In eq.~(1), the summation goes over $N$ effective
sites which can in general represent mers of an arbitrary
semiconducting like molecular chain.  The coupling of the molecule to
the leads will be described using energy independent parameters,
$\Gamma_a$, $\Gamma_b$.\cite{Kostyrko-02} The effect of the potential
ramp related to the source\---drain voltage will be simulated by a
site dependent shift of the on-site energy parameters, $E_a$ and
$E_b$. In the same way, the effect of the local probe will be taken
into account by adding a site energy shift $E_L$ at a site
$L$. Finally, to account for the energetic inequivalence of the
bonding atoms we use a site energy shift $E_S$ at the first and the
last site of the chain. Using transfer matrices
  \cite{cjl1,cjl2}, the transmission coefficient for the above
  Hamiltonian can be easily computed for any finite system. The
energy separation between the bands should be large enough to describe
a molecular chain with semiconducting properties. When contacted to
the leads the molecule will form a junction with the conductance
decreasing exponentially with the increase of the chain length, as
discussed in many papers (see, e.g.  Ref.\onlinecite{Joachim-02}). The
presence of the local shift of the potential can bring about the
appearance of a localized state in the gap of the system. If the
molecule is strongly coupled to the leads and the in-gap state has
large enough spatial extension it can couple directly to electronic
states of the leads and participate in the resonant transport through
the molecule, as described in the previous section.

%
%
The orbitals of the peripheral atoms are weakly coupled by the
effective hopping $t_\mathrm{eff}$, which could be computed by
eliminating all the other orbitals in the chain. Approximately, the
whole molecule can be described in a limited energy range as a
two\---site system with a weak coupling $t_\mathrm{eff}$ between the
sites. Note, that these states are located mostly on the peripheral
sites and are almost symmetric to each other. The most interesting
feature here is the probe-induced level, which is located close to the
probe. This state can be interpreted as a localized state split off
from the antibonding band by the strong attractive local potential
created by the probe.

%
%

 The procedure of reproducing the {\it ab initio} results with
  the model system is facilitated by fact that the
  Hamiltonian~(\ref{H_M}) is exactly solvable for any number of sites
  and parameter values and its eigenvalues are give by:

\begin{eqnarray}
\varepsilon_{\nu}^{(\pm)}
&=&
\frac{1}{2}
\left[
E_a+E_b+2(t_a+t_b)\,\cos\left(\frac{\nu\pi}{N+1}\right)\right]
\nonumber \\[0.5em]
&\pm&\frac{1}{2}
\sqrt{
\left[
E_a-E_b+2(t_a-t_b)\,\mathrm{C}\right]^2
+
16\,t_{ab}^2\,\mathrm{C}^2
}
\end{eqnarray}

\noindent where $\mathrm{C}=\cos\left(\frac{\nu\pi}{N+1}\right)$.

\begin{figure}[h]
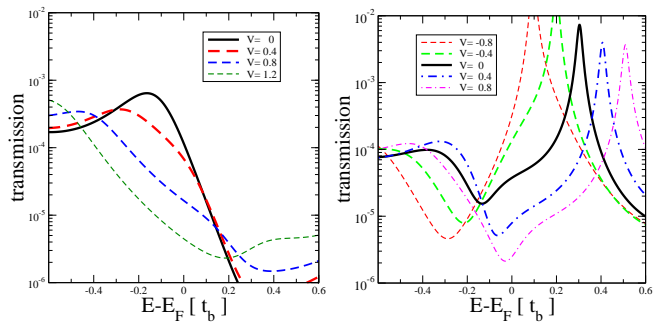

\includegraphics*[width=0.49\columnwidth]{fig6_r.eps}
\includegraphics*[width=0.49\columnwidth]{fig7_r.eps}
\caption{\label{T_semic1d}(color online) Evolution of the transmission
  in the vicinity of the Fermi level ($E_{\mathrm F}=0$) with the bias
  voltage for the molecular chain of 8 mers, described by the
  Hamiltonian~${\cal{}H}_M$ coupled to the leads with $\Gamma_{a,b}=
  0.2$. The used model parameters are: $E_a= 3$, $E_b= -3$,
    $t_b=-t_a= 1$, $t_{ab}=0.11$, $L=2$, $E_S= 2.5$. Left:
  transmission for the system without the probe; right:
  transmission for the system with the probe corresponding to
  $E_L=-2.1$ located near the $L=2$ atom.}
\end{figure}

We start from setting the energy scale of our parameter unit $t_b$ to
a value of 4~eV which (as to the order of magnitude) is suggested by a
hopping integrals often used to describe the hydrocarbon
systems.\cite{Horsfield-96} For simplicity we also put $t_b=-t_a$ and
treat in what follows $t_{ab}$ as a small number. The energy gap of
the spectrum of Hamiltonian~(\ref{H_M}) is given, for large $N$, by:
$E_g=\sqrt{\left(|E_a-E_b|-2|t_a-t_b|\right)^2+ 16\,t_{ab}^2}$.  On
the basis of our {\sc siesta} and {\sc smeagol} computations the
energy gap between the main parts of energy spectrum of the ODT
molecule coupled to the leads is given by $\sim{}8$~eV.  The last
value can be well reproduced (for small $t_{ab}$) by taking
$E_a=-E_b=3t_b$. In order to fit the position of the sulfur derived
HOMO band, as well as the shape of the 2~eV wide transmission peak
just below $E_F$ (cf. figure~\ref{transmission_ODT}) we are left with
3 more parameters: $E_S$, $\Gamma_a=\Gamma_b$ and $t_{ab}$.  Adjusting
their values with a trial and error method to the {\it ab initio}
results we get: $\Gamma_a=\Gamma_b=0.2t_b$, $E_S=2.5t_b$ and
$t_{ab}=0.11t_b$. Obtained in this way transmission function near
$E_F$ is presented in figure~\ref{T_semic1d} (left panel). We conclude
that it compares fairly well to {\it ab initio} results shown in
figure~\ref{transmission_ODT} (left upper panel) taking into account
the very approximate nature of the two\---orbital model as well as all
the approximations implied by the one\---particle approach.

The plots clearly possess the same features as the transmission plots
computed using DFT. In the absence of the charge probe, the
transmission is dominated by a very wide peak which comes from the two
levels HOMO-1 and HOMO, located mainly on the both ends of the
chain. Since the effective hopping between the peripheral sites for
this 8-site chain is much smaller than the applied intermediate values
of the molecule\---lead couplings, the transmission is much less than
unity in the center of the peak. Suppression of the peak by the
  probe is due to the non\---equivalent coupling of the HOMO states to
  the leads, resulting from the asymmetric redistribution of the
  relevant wave functions.

In order to quantitatively simulate the effect of the external
  probe we shift the local energy of the bonding and antibonding
  levels by a common value $E_L=-2.1t_b$ which reproduces the position
  of the peak from {\it ab initio} computations for the $+2|e|$ probe.
  The height of the transmission peak is then more or less the same as
  the one obtained for the ODT junction
  (cf. figure~\ref{transmission_ODT}, middle column), what
  independently validates our choice of $E_L$. At the same time the
  wide peak below $E_F$ is suppressed in comparison to the case
  without the probe (cf. figure~\ref{transmission_ODT}, lower left
  panel), although the value of transmission at the local maximum is
  by a factor of 3 lower than the one of the {\it ab initio} results.

The evolution of the transmission with the bias voltage is also
similar to the one observed in the DFT computation: the height of the
wide peak decreases and it begins to separate into two peaks
corresponding to the two orbitals, with one located mostly on the
source lead and the other one mostly on the drain lead.  The position
of the probe\---induced transmission peak corresponds rather
accurately to the energy of the quasi\---localized state which
  can be also found in PDOS (not shown here). The state couples much
better to the left electrode and because of this reason its position
follows rather closely the electrochemical potential of the left
lead. This asymmetry of the coupling also explains the reduced
  value of the transmission for the peak, which is considerably
  smaller than unity.

\begin{figure}[h]
\includegraphics*[width=\columnwidth]{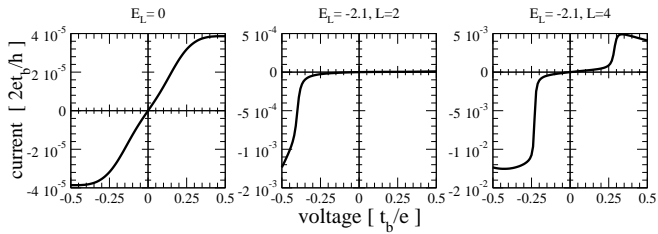}
\caption{\label{IV_semic1d} $I-V$ dependence of the model system
  described by the Hamiltonian ${\cal{}H}_M$ for the same parameters
  as in figure~\protect{\ref{T_semic1d}}.}
\end{figure}

The $I-V$ dependence (see figure~\ref{IV_semic1d}) corresponding to
the transmission data from figure~\ref{T_semic1d} also compare well
with the {\it ab initio} results from figure~\ref{current_ODT}.
The current for $E_L=0$ shows the same smooth voltage dependence as
the corresponding DFT results in the absence of the probe, with an
almost linear dependence in the region of the small voltage and a
saturation slightly above 0.25~$|t_b|\sim{}1$~V. The current for
$E_L=-2.1t_b$ and $L=2$ is also qualitatively similar to the DFT
results in the presence of the probe, with a sharp rise of the
absolute current for voltages near -1~V. In the present case, however,
the rectification ratio is significantly larger than in the DFT
results, amounting to about 50. Another difference with the DFT
  results is found for the central location of the probe, where the
  model calculations show a gap\---like shape in the current in the
  region $-0.25t_b<eV<0.25t_b$. This is due to departure of the
  transmission peak from $E_F$ at $V=0$ in the model calculation for
  $L=4$. We expect that some of the discrepancies can be removed by a
  selfconsistent treatment of the model, extended to include
  explicitly electron interactions which can modify the positions of
  the molecular levels with respect to the Fermi levels of the leads.

\section{Conclusion}
In conclusion, we have shown that the presence of a local charge in
the vicinity of a molecular backbone can lead to significant
rectification in the otherwise symmetric molecule.  The external local
potential is shown to split a quasi\---localized state from the
unoccupied band. The state is centered near the probe position and
significantly contributes to the electron transport when the voltage
increases beyond a threshold value. The rectification is due to the
strong energy dependence of the quasilocalized state on the
polarization of the bias.  The resulting rectification ratio depends
non\---linearly on the value of the potential, and the substantial I-V
asymmetry is observed only for strong enough potential values. We
showed that the application of the external potential does not
necessarily imply a current reduction, in fact we observed a increase of
the current value as compared to the symmetric case when the stronger charge
probe was included. The latter finding qualitatively agrees with
the results of a recent experimental work.\cite{Diez-Perez-09}

The proposed parametric model of the molecular junction in the field
produced by the charge probe can be considered as a generic minimal
model of a semiconducting\---like linear molecule with an asymmetry
induced by a local potential. The successful explanation of the gross
features of the DFT results, with a simple choice of parameter values,
suggests that the described mechanism of the rectifying behaviour can
be a rather common feature of such junctions. The main requirement for
the device to be useful for applications is that the charge probe has
to be strong enough to generate a quasilocalized level within the
HOMO-LUMO energy gap, close enough to the HOMO level. In this way one
can generate a quasilocalized state with a substantial spatial overlap
with the electronic states of the both leads, so that it is active in
transport for experimentally accessible voltage values.
\section*{}
\acknowledgments
This work was supported by Ministry of Science and Higher Education
(Poland) from sources for science in the years 2009–2012 within a
research project. VMGS thanks the Spanish Ministerio de Ciencia e
Innovaci\'on for a Juan de la Cierva fellowship and the Marie Curie
European ITNs FUNMOLS and NANOCTM for funding.


\begin{thebibliography}{12}
\bibitem{siesta-a}
Ref.\protect{\onlinecite{siesta}} and the {\sc siesta}
manual give detailed explanations of the meaning of the
simulation parameters used in our work.
\bibitem{convergence}
We were unable to obtain convergence for higher voltage ranges for
some of the data presented in figures~\protect{\ref{transmission_ODT}}
and \protect{\ref{current_ODT}}.
\end{thebibliography}
\end{document}